# Characterizing patterns in police stops by race in Minneapolis from 2016-2021

Onookome-Okome, Tuviere[a,*]; Gorondensky, Jonah[a]; Rose, Eric[a]; Sauer, Jeffery[b]; Lum, Kristian[c]; Moodie, Erica E.M.[a]

*: corresponding author
a: McGill University, Department of Epidemiology, Biostatistics and Occupational Health
b: University of Maryland College Park, Department of Geographical Sciences
c: Twitter, San Francisco

## Abstract

The murder of George Floyd centered Minneapolis, Minnesota, in conversations on racial injustice in the US. We leverage open data from the Minneapolis Police Department to analyze individual, geographic, and temporal patterns in more than 170,000 police stops since 2016. We evaluate person and vehicle searches at the individual level by race using generalized estimating equations with neighborhood clustering, directly addressing neighborhood differences in police activity. Minneapolis exhibits clear patterns of disproportionate policing by race, wherein Black people are searched at higher rates compared to White people. Temporal visualizations indicate that police stops declined following the murder of George Floyd. This analysis provides contemporary evidence on the state of policing for a major metropolitan area in the United States.

# 1. Introduction

Instances of police brutality have, once again, surged to the forefront of public consciousness. In particular, the murder of George Floyd on May 25, 2020, by on-duty members of the Minneapolis Police prompted widespread protests, public discourse, and calls for the abolition of the police in the United States and around the globe. Yet the murder of George Floyd is not an isolated incident of police brutality in Minneapolis nor Minnesota more broadly. On July 6, 2016, Philando Castile was fatally shot in the neighbourhood of Falcon Heights, a residential area in the Minneapolis-St. Paul metropolitan area. Responding in 2016, former governor Mark Dayton stated "would this [fatal shooting] have happened if those passengers, the driver and the passengers, were white? I don't think it would have." (Pugmire & Bakst, 2016). Although these events have received immense social and political attention[1], troubling police interactions have continued in and around Minneapolis. While preparing this article, another individual, Daunte Wright, was shot and killed at the hands of police in Brooklyn Park, Minnesota, on April 11, 2021. These instances of extreme police brutality are modern examples in a legacy of violence against Black, Indigenous, and People of Color (BIPOC) extensively documented in Minnesota (Anderson, 2019; Farley et al., 2011; Pfeifer & Bessler, 2004).

The greater Minneapolis-St. Paul metropolitan area is home to approximately 3.4 million people. Minneapolis serves as Minnesota's largest city with approximately 420,000 residents as of 2019 (U.S. Census Bureau, 2019). 63.6% of Minneapolis residents are White, with Black residents accounting for 19.2% of the population (U.S. Census Bureau, 2019). Community-based reports submit that although Black Minneapolis residents are a much smaller portion of the population,

---

[1] In addition to George Floyd and Philando Castile, several police-involved shootings in the Minneapolis area have drawn intense media coverage and controversy. Recent examples include the shooting of Jamar Clark, Justine Damond, and Thurman Blevins to name just a few incidents from Minneapolis alone.

Black residents have more interaction with the police and more instances of police violence compared to White residents (MPD150, 2020). Following the murder of George Floyd, a number of large public institutions distanced themselves from the Minneapolis Police Department (MPD), with notable examples including the Minneapolis Public Schools terminating an existing MPD contract and the University of Minnesota pledging to limit further MPD contracts (Faircloth, 2020; Strauss, 2020).

Community-based organizations in Minneapolis argue acts of public separation do not go far enough, calling for systematic evaluations on the interactions of MPD and Minneapolis residents (MPD150, 2020). These calls are made in explicit recognition of the numerous forces governing the dynamics of police interactions in the United States. Black people are disproportionately the victims of police brutality in the United States, and police brutality is often excused or ignored by due to pervasive racism (Chaney & Robertson, 2013). An additional and important dynamic is that negative police interactions, including police brutality, are rarely punished, leading some legal scholars to argue that 'police brutality has been effectively decriminalized in this country [the United States]' (Davis, 1994). Although instances of accountability for police brutality exist, several recognized obstacles from both within the institution of police, usually recognized as the 'Blue Wall of Silence' or 'Blue Code of Silence', and society at large, such as the response of 'Blue Lives Matter' to 'Black Lives Matter', pose serious obstacles to addressing negative consequences of racial bias in police actions (Cooper, 2020; Skolnick, 2002; Solomon et al., 2021).

There have been some studies examining policing and racial bias in the greater Minneapolis area. Examples including research by Gorsuch and Rho (2019), Ritter (2017), Ritter and Bael (2009), Wexler (2020), as well as related work on racial bias in the Minnesota justice system by Martin and Thompson (2001) (Martin & Thompson, 2001; Ritter & Bael, 2009; Ritter, 2017; Gorsuch & Rho, 2019; Wexler, 2020). Gorsuch and Rho, using the same publicly available MPD stop data we describe later in the paper, found that Indigenous women were searched and arrested at a rate of more than double that of women belonging to any other racial group (Gorsuch & Rho, 2019).[2] Moreover, Gorsuch and Rho use descriptive mapping to highlight the concentration of policing in neighbourhoods with a high Indigenous population (Gorsuch & Rho, 2019). Ritter (2017) investigated the role of race in MPD stop and search data from 2002, finding evidence of the effect of race on police stop decisions, but not police on search decisions (Ritter, 2017). Earlier work by Ritter and Bael (2009) used the same data from 2002 and found evidence that Black drivers were stopped 5.5% less often at night compared to during the day, supporting the "veil of darkness" theory which states that racial profiling is less likely when visibility is obscured (Ritter & Bael, 2009). Investigating race visibility and police stops, Wexler (2020) demonstrated similar findings to Ritter and Bael (2009), although to a lesser extent (Wexler, 2020).

While studies on policing and racial bias in Minnesota may be limited in number, numerous studies on policing and racial bias have been published from locales across the United States over the past several decades. Early work by Harris demonstrated that Black individuals, especially Black individuals of driving age, were more than twice as likely to get a ticket compared to White drivers across several major police departments (Akron, Toledo, Dayton, and Columbus/Franklin County) in Ohio (Harris, 1999). Nationwide audits of police-public contact from the same time period found that although there were far fewer Black drivers compared to white drivers in the United States, a greater proportion of Black drivers were stopped (12.3% of Black drivers compared to 10.4% of White drivers) (Langan et al., 2001). Later studies on police stops in Boston showed that Non-white individuals are stopped and arrested more often than White individuals, even after adjusting for potentially confounding social and criminological factors (Fagan et al., 2015, 2016). In New York City, individuals of African and Hispanic descent tend to be stopped approximately twice as often as Whites for suspected violent crimes and weapons offenses (Gelman et al., 2007). Gelman et al. also found evidence of high stop rates for minorities in predominantly White precincts (as well as high stop rates for Whites in predominantly minority precincts) and argue that such "racial incongruity" stops indicate the police act on "out of place" suspicions (Gould & Mastrofski, 2004; Alpert et al., 2005; Gelman et al., 2007). This latter finding aligns with research by Roh and Robinson (2009) who demonstrate the utility of analyzing stop data at both the micro- (e.g., individual) and macro- (e.g., areal units) level when investigating racial disparities in traffic stops (Roh & Robinson, 2009). Another study situated in Durham, North Carolina, demonstrated a positive association between daytime stops and an individual being a Black male even after accounting for officer, rather than neighborhood, factors (Taniguchi et al., 2017). Buerger and Farrell provide a summary of the notable litigation surrounding racial profiling in the late 1990s and early 2000s, arguing that narrowing legal definitions versus widening public definitions limited system-wide solutions, even in the face of increasingly substantiated demonstration of racial disparities in stop rates across the United States (Buerger & Farrell, 2002).[3] Studies with a focus outside the United States have also found racial bias in policing. The notable and early 'Black and Blue' study of police stops in London, England, showed that Black individuals were 2.8 times more likely to be stopped compared to their representation in the underlying population, even as Black and White individuals had equally mild reactions during police contact (Norris et al., 1992). A systematic review of international, empirical studies examining racial prejudice and police stops found that Black men were the most commonly stopped racial group amongst studies situated in the United States, England, Wales, and Netherlands (Ara et al., 2021).

---

[2] We use the term Indigenous and Indigenous people(s) throughout the manuscript as this recognizes the independence and original inhabitance of Indigenous peoples throughout the North American continent. Indigenous peoples are coded as 'Native American' in the Minneapolis Police Stop dataset.

---

[3] Readers interested in more extensive literature reviews on racial bias and disparities in traffic stops should look to both Roh and Robinson (see section 'Literature Review') and Gorsuch and Rho 2019 (see section 'Background - Disparities in Police Stops and Their Effects') (Gorsuch & Rho, 2019; Roh & Robinson, 2009).

Many of the extant studies on racial bias in policing have made use of limited access data. The expansion of open data in the last decade has led to new avenues for research into police interactions. Recently, Pierson et al. (2020) collated data on approximately 100 million traffic stops from across the United States, providing robust evidence on several factors related to racialized policing including support for the "veil of darkness" theory, excess stops of Black and Hispanic drivers compared to White drivers, and policy changes like the legalization of recreational marijuana failing to address racial disparities in police stops (Pierson et al., 2020).Smaller-scale studies also using open data have found similar effects. A study using nearly a decade of publicly available traffic-stop records from Louisville, Kentucky found that Black drivers were more likely to be stopped even when compared with other stops of similar characteristics (Vito et al., 2020). Another analysis using open data from the Washington, D.C., metropolitan police department found that Black individuals comprised nearly 90% of all stops even though the Black population within the jurisdiction was only 51% of the population, and that Black individuals had a disproportionately higher arrest rate compared to the White population in an area (Fielding-Miller et al., 2016). Open data are one of the many emergent tools - alongside others including but not limited to public video recording, police body cameras, apps that automatically report potential incidents of misconduct, social media, and more - that researchers, policy makers, and activists have identified as tools that may help hold police accountable (S. E. Walker & Archbold, 2018).

Given the recent events in Minneapolis and the calls from community-based organizations to evaluate police interaction, this study set out to quantitatively assess the relationship between race and police interaction in Minneapolis, Minnesota with a particular focus on the period around the murder of George Floyd. This study makes use of a large open dataset provided by the Minneapolis Police Department that includes more than 170,000 police stops, representing a major opportunity that has been largely unavailable in previous studies of police interaction. In addition, we implement a recent computational advance in generalized estimating equations to handle large datasets with varying cluster sizes, which is important given differences in police activity by neighbourhood. Using these data, we provide highly detailed descriptive information on police stops that occurred in Minneapolis from November 2016, to March 2020, both at the individual- and the Census Tract-level, illustrating how police stops disproportionately impact those who are BIPOC. We then quantitatively assess the extent to which race is associated with vehicle and persons searches among those who are stopped, finding further evidence of racial bias in police interactions. Lastly, we highlight sudden and drastic temporal variation in police interaction following the murder of George Floyd.

## 2. Methods

### 2.1 Data

MPD stop data were collected from the Open Minneapolis data portal (https://opendata.minneapolismn.gov/) (City of Minneapolis, 2021). A data dictionary was acquired from the City of Minneapolis Geographic Information System (GIS) Office. The police stop data contain records of all police interventions occurring in the city of Minneapolis from late-2016 to the present day. For the present study we used police stop data ranging from October 31, 2016, to March 15, 2021. The data contain variables describing stop information and incident identification numbers, information on the stopped individual (e.g., pre-stop information on suspect race communicated by dispatch, reported race, and reported gender), action resulting from the stop (e.g., person search, vehicle search, booking, no action, etc.), and geographic information on the stop (e.g., latitude and longitude, Minneapolis neighbourhood, and Minneapolis police precinct). The primary outcomes of interest are stops resulting in a person search or a vehicle search, both of which are available in the police stop data. These data originate from MPD police officers conducting stops within the metropolitan boundaries of Minneapolis, and the resulting data are made available by the City of Minneapolis government.

A few variables, variable transformations, and data cleaning steps require elaboration. Time of stop was converted from a continuous variable to a discrete variable indicating if the stop took place during dusk or sundown using the suncalc package in R (Thieurmel & Elmarhraoui, 2019). Additionally, reported gender contained four categories: male, female, gender non-conforming, and unknown. Although gender non-conforming was reported in very low numbers, we opted to retain it as a category rather than collapse groups (Ansara & Hegarty, 2014). We removed stops in which the reason for the stop was truancy or curfew violations, as these reasons were reported in less than 0.01% of all stops. Lastly, we categorized individuals reporting East African for race as Black. The East African race category is available in the police stop data due to the sizable population of Somali and Somali Americans living in Minneapolis.

## 2.2 Descriptive Analysis and Mapping

Several descriptive data analysis techniques and visualizations were used to characterize the distribution of police stops in Minneapolis. Racial demographics for the city of Minneapolis were extracted from the US Census. The racial distribution of the city population was compared to the racial distribution of police stops to assess whether police stops by race are proportionally distributed. We then provide summary statistics of the police stop data along several dimensions including pre-stop information on suspect race communicated by dispatch, reported race, time of day, recorded problem, and reported gender. We describe the data using both descriptors of race (pre- and during stop) as reported race reflects post-event information, whereas initial information on race is reported by dispatch when available.

Following the descriptive tabulations, we generate several data visualizations. To begin, we graph the absolute number of police stops by race and the percentage of police stops resulting in a person or vehicle search by race. These initial visualizations consider all police stops across the study period without regard to possible temporal or geographic trends. We further characterized the data using time series graphs and thematic maps. The time series graphs describe the racial distribution of police interactions by race for each month-year across the study period. The thematic maps show police stop rates across Minneapolis at the Census Tract level. Specifically, these maps show police stop rates for White and Black race groups for each complete year in the study period (e.g., 2017, 2018, and 2019). These additional visualizations are used to describe temporal or geographic patterns present in the police stop data.

## 2.3 Regression Analyses

We sought to evaluate the relationship between race and being searched by the MPD while accounting for potential confounding factors and to characterize any changes in MPD practices in the wake of George Floyd's murder and subsequent protests. To account for the correlated nature of police stops within a given geographical area, generalized estimating equations (GEE) with a logistic link were employed separately for both outcomes (person search and vehicle search) (Liang & Zeger, 1986) to estimate how police stop behaviours vary among Black people compared to White people across Minneapolis.

Two different models were fit for each outcome type. The first involved specifying univariate models that modeled search (the outcome) as a function of race, whether the stop occurred after the murder of George Floyd, and their interaction as the only independent variables. Let $Y_{ij}$ denote the binary outcome (person search or vehicle search) and let $X_{ij}$ denote a vector of indicator variables indicating the race of the individual stopped at stop $j$ in neighbourhood $i$ for $i = 1, \dots, N, j = 1, \dots, n_i$. Let $A_{ij}$ denote an indicator variable for whether the police stop occurred after the murder of George Floyd. The marginal mean model for $Y_{ij}$ was then given by

$$E(Y_{ij}|X_{ij}; \beta) = \mu_{ij}(\beta) = g(\beta_0 + X_{ij}^T\beta_1 + A_{ij}\beta_2 + A_{ij}X_{ij}^T\beta_3)$$

such that $g(\cdot)$ denotes the logit link function. We will denote the conditional variance of $Y_{ij}$ by $v(\mu_{ij})$ and therefore $v(\mu_{ij}) = Var(Y_{ij}|X_{ij}; \beta) = \mu_{ij}(1 - \mu_{ij})$. We assume an exchangeable correlation structure, so we have that $Cor(Y_{ij}, Y_{ik}) = \rho$. Let $Y_i = (Y_{i1}, \dots, Y_{in_i})$ denote the vector of outcomes for the $n_i$ stops in neighborhood $i$. The correlation matrix of $Y_i$ is then given by $R_i(\rho) = \rho J_{n_i} + (1 - \rho)I_{n_i}$ where $I_{n_i}$ is an $n_i \times n_i$ identity matrix and $J_{n_i}$ is an $n_i \times n_i$ matrix of ones. The covariance matrix of $Y_i$ which we will denote by $V_i(\rho, \beta)$ is given by $V_i(\rho, \beta) = B_i^{1/2}(\beta)R_i(\rho)B_i^{1/2}(\beta)$ where $B_i(\beta) = \text{diag}\{v(\mu_{i1}), \dots, v(\mu_{in_i})\}$. The GEE then has the form

$$U(\rho, \beta) = \sum_{i=1}^{N} D_i^T(\beta) V_i^{-1}(\rho, \beta)\{Y_i - \mu_i(\beta)\}$$

such that $D_i^T(\beta) = \partial\mu_i(\beta)/\partial\beta$.

The second involved specifying multivariate models that include several factors the police may use when deciding to stop and search an individual. In addition to the main effect of interest (race of the stopped individual), the models also accounted for the reported gender of the stopped individual, the pre-race of the stopped individual, the reported problem prompting the stomp, the ultimately determined reason for the stop, an indicator for whether the stopped occurred after sunset and the latitude and longitude of the police stop. Let $Z_{ij}$ denote a vector of these covariates at stop $j$ in neighbourhood $i$. The marginal mean model for $Y_{ij}$ for the multivariate model was then given by

$$E(Y_{ij}|X_{ij}; \beta) = \mu_{ij}(\beta) = g(\beta_0 + X_{ij}^T\beta_1 + A_{ij}\beta_2 + A_{ij}X_{ij}^T\beta_3 + Z_{ij}^T\beta_4).$$

The GEE for the multivariate model then had the same general form as the univariate model with only the marginal mean model changed. Given the number of associations considered, we opted to control the false discovery rate at 0.05 using the Benjamini-Hochberg procedure (Benjamini & Hochberg, 1995).

**2.4 Sensitivity Analyses**

Information on the race of the individual stopped by police is centrally important to the present research. However, police stop data suffers from missingness and unclear categories that can hinder examinations by race (Alwd 6th Ed. Richardson et al., 2019; Webster, 2021). A large proportion of MPD stop data describes the race of the stopped individual as Unknown. To check the sensitivity of our results against scenarios where the race category of Unknown is systematically assigned to one race, we repeat the regression analyses assuming different proportions of race for those identified as Unknown. We randomly select those identified as Unknown and assign them either a white or Black race in different proportions including: 0% White, 100% Black; 20% White, 80% Black; 40% White, 60% Black; 60% White; 40% Black; 80% White; 20% Black; and 100% White, 0% Black. Examining these scenarios allows us to better understand how regression estimates might change if more information was available for individuals identified as having an Unknown race. In particular, the most extreme of these imputation approaches allows us to investigate the extent to which our results might be changed by data that are systematically incorrectly reported as Unknown.

## 3. Results

### 3.1 Descriptive Analysis and Mapping

Between October 31$^{st}$, 2016, and March 15$^{th}$, 2021, there were 173,290 police stops in the city of Minneapolis. Table 1 shows that while Black individuals make up only 19.2% of the Minneapolis population at large, Black individuals make up 34.7% of all police stops. Comparatively, white people make up 63.6% of the Minneapolis population yet only 20.9% of all police stops. At a very crude level, this comparison indicates a differential relationship in police interaction by race in Minneapolis.

Summary statistics of the police stop data by pre-race, reported race, recorded problem, and reported gender are presented in Table 2. There were 19,224 (11.1%) person searches and 12,067 (7.0%) vehicle searches among the analyzed stops. Among stops with race information provided by dispatch, Black people made up the largest share of person searches (n=6267, 32.6%) and vehicle searches (n=2485, 20.6%). This share greatly increased once the race of the stopped individual was determined – Black people made up 65.6% of all person searches and 66.4% of all vehicle searches. The most frequently reported problem reason amongst person searches was for 'Traffic Law Enforcement' (n=7519, 39.1%) and 'Suspicious Person' (n=7304, 38.0%), whereas 'Traffic Law Enforcement' was cited in the majority of vehicle searches (n=7593, 62.9%). There are far more males stopped compared to females, regardless of search.

Figure 1 Panel A demonstrates that, in absolute numbers, there are far more stops involving Black individuals than individuals of any other race, although as seen in Table 1, there are also a substantial number of stops involving individuals of unknown race. Panel B shows the number of searches *relative* to the number of stops within that given race. While Black individuals are searched the most in absolute terms, it is Indigenous individuals who are searched most frequently when accounting for the number of stops. Indigenous people are searched nearly 30% of the time when stopped, with Black people are searched approximately 25% of the time, with decreasing percentages across the rest of the categories. Interestingly, Figure 1 Panel B shows that when the race is recorded "unknown" there are relatively few searches.

A time-series plot of MPD stops by race is shown in Figure 2. The figure illustrates several important temporal trends in the MPD stop data. While police stops have slightly decreased from 2016 to 2020, the distribution of stops by race has remained consistent. Between November 2016 and March 2020, Black people made up between 28.5% to 43.5% of stops in a given month-year, whereas White people made up between 16.4% to 25.6% of stops in a given month-year. Police stops tend to increase and peak between March and August, with the lowest number of stops tending to occur in December and February.

The time-series plot also contains annotations of two major events in the Minneapolis area, specifically implementation of COVID-19 restrictions and the murder of George Floyd. The first case of COVID-19 in Minnesota was recorded March 6th, 2020, with stay-at-home orders issued on March 27th, 2020. These initial months of the COVID-19 pandemic saw similar levels of police stops with 3096 stops in March 2020, 3018 stops in April 2020, and 3192 stops in May 2020. However, the months immediately following George Floyd's death (May 25th, 2020) saw a sharp decline in the total number of police stops and, most importantly, apparent changes in who was stopped. For example, Black people made up 42.9% of stops in May 2020, yet dropped to 18.5% of stops in June 2020. Notably, stops with a reported unknown race increased from 33.3% in May 2020 to 66.5% in June 2020.

The final set of visualizations describe geographic patterns of person search stop rates across Minneapolis at the Census Track (CT)-level (Figure 3). These maps depict clear differences in both the location and intensity of police activity between Black and White people. The average CT-level person search stop rate across the study period was 5.99 (95%CI 5.91-6.07) per 100 Black people. In contrast, the average CT-level person search stop rate was 0.57 (95%CI 0.57-0.58) per 100 White people. Areas of Minneapolis with the highest CT-level person search stop rates include East Isles (CT ID 1066, person search rate 140 per 100 Black people), Downtown West (CT ID 1044, person search rate 90.7 per 100 Black people), Camden Hawthorne (CT ID 22, person search rate 60.3 per 100 Black people), and East Calhoun (CT ID 1080, person search rate 59.4 per 100 Black people). Apart from Camden Hawthorne, these areas are largely high-income, have a large share of White residents, and are in West/Southwest Minneapolis. Comparatively, CTs with the highest person search stop rates for White people are North North-Hawthrone (CT ID 1023, person search rate 8.8 per 100 White people), Camden Hawthrone (CT ID 22, person search rate 6.6 per 100 White people), and Near North (CT ID 1029, person search rate 3.5 per 100 White people). These areas are all located in Northwest Minneapolis, an area with a larger Black residential population. The area of Camden Hawthorne offers a compelling contrast. Although Camden Hawthorne has some of the highest person search rates, there is a striking difference in the magnitude of the rate between Black and White people.

### 3.2 Regression Analyses: Association Between Race and Police Interaction

Univariate and multivariate logistic regression using generalized estimating equations (GEEs) were used to model the relationship between race and police searches to account for clustering at the neighbourhood level. Table 3 displays odds ratio for both the univariate and multivariate models across both outcomes before the murder of George Floyd. Looking first at person search, we see that, as compared to White people, the crude odds ratio (OR) was 2.01 (95% confidence interval, CI, 1.59-2.53), indicating an increased odds of being searched among Black people as compared to White people. In the multivariate model for person search, the odds ratio decreased slightly to 1.72 (95% CI 1.54-1.92), still indicating a significantly increased odds of Black people being searched during a police stop as compared to White people.

The results were similar when looking at whether an individual's vehicle was searched, though the odds ratios were slightly larger. Black people had an odds ratio of 2.26 (95% CI 2.08-2.45) in the univariate model and 2.02 (95% CI 1.83-2.24) in the multivariate model. These numbers imply that, when stopped by the police, Black people have approximately twice the odds of having their vehicle searched as compared to White people.

We found that Indigenous people had the largest odds of being subject to a person and vehicle search for all models except for the univariate model for vehicle search. For the univariate model, the odds ratio was 3.96 (95% CI, 3.35-4.68) for person search and 2.21 (95% CI, 1.88-2.60) for vehicle search. In the multivariate model, the odds ratio was 3.52 (95% CI, 3.01-4.11) for person search and 5.36 (95% CI, 4.37-6.56) for vehicle search.

Table 4 displays odds ratio for both the univariate and multivariate models across both outcomes after the murder of George Floyd when compared to White people after his murder. We found that the odds ratios for Black people increased after the murder of George Floyd compared to before. For the person search outcome, the crude odds ratio (OR) was 2.77 (95% CI, 2.07-3.71) for the univariate model and 2.37 (95% CI, 1.94-2.09) for the multivariate model. For vehicle searches, the crude odds ratio (OR) was 3.11 (95% CI, 2.45-3.95) for the univariate model and 2.55 (95% CI, 1.92-3.37) for the multivariate model.

### 3.3 Sensitivity Analyses: Unknown Race

In sensitivity analyses, we study the impact of misclassification of individuals' race as Unknown under different assumptions regarding the true, unrecorded race. Figure 4 shows the estimated odds ratios and 95% confidence intervals for the estimated association between Black and a person/vehicle search using different proportions of White-Black random racial replacement. Under the extreme, and implausible, assumption that all those classified as Unknown were in fact White, some of the estimated effects in univariate models were found not to differ significantly from the null hypothesis of no impact of race (Figure 4a, 4b, and 4d). In contrast, under all settings considered for the true race of those classified as Unknown, *all* the adjusted models support the conclusion that Black individuals have higher odds of person and vehicle searches both before and after the murder of George Floyd. These findings indicate the robustness of our initial results.

**4. Discussion**

The results of the present analysis allow for several observations regarding race and policing in Minneapolis. Firstly, we can answer our original question of whether race is associated with being searched by police in Minneapolis. This question has been addressed many times in both lay press and academic research, and our analysis contributes to the ongoing dialogue by using open-source policing data and methods that can account for confounding and correlated data. Crude risks calculated both in the press and in this paper suggest strongly that Black people are more likely to be searched by the police than White people. Our multivariate analyses allow us to take this statement further. Our results tell us that Black people are statistically significantly more likely to have both their person or vehicle searched in Minneapolis when stopped by police as compared to White people even after accounting for several potential confounds including time of day, the problem and the reason for the stop, before and after the murder of George Floyd, and accounting for clustering within neighbourhood and more. Though the effect size slightly dampens between the crude and adjusted models, Black people clearly have increased odds of being searched.

Temporal visualization of police stops by race revealed several interesting findings. We observed a sharp decrease in the total number of police stops in the month immediately following the killing of George Floyd. This finding is closely related to the disputed 'Ferguson Effect' (Davey & Smith, 2015; Gross & Mann, 2017; Mac Donald, 2015; Pyrooz et al., 2016; Rosenfeld et al., 2015), which suggests that consistent, year-over-year declines in the rate of the violent crime since the 1990s began to reverse in the mid-2010s following massive public demonstrations over the 2014 shooting of Michael Brown in Ferguson, Missouri (Mac Donald, 2015). The implied mechanism underlying the Ferguson Effect is that police disengage from regular behavior following public outcry over their actions (e.g., media coverage of excessive use of force), ultimately resulting in more opportunity for crime (Mac Donald, 2015). Thus, the Ferguson Effect has two key components: (1) a decrease in police activity and (2) an increase in crime. Empirical analyses examining recent crime rates are generally limited due to a lack of data (Rosenfeld, 2016; Rosenfeld & Fox, 2019). A large sample of monthly criminal offense data from 81 large US cities found no evidence to support change in overall, violent, and property crime rates, even if select cities experienced increases in homicide in 2014 (Pyrooz et al., 2016). Similarly, a study using data from 53 large US cities covering the years 2010 to 2015 found no relationship between declines in arrests and homicide rates (Rosenfeld & Wallman, 2019). However, other studies have substantiated a decrease in police activity (also called 'de-policing') following instances of major public outcry (Shjarback et al., 2017). For example, a study of police agencies across Missouri found that agencies carried out nearly 70,000 fewer traffic stops in the year following Michael Brown's shooting, and that this effect was especially pronounced in agencies with a higher proportion of African-American residents (Shjarback et al., 2017). The suggested reasons behind de-policing are varied, including discouragement among police managers (Nix & Wolfe, 2018), police officer concerns over public and judicial repercussions (Oliver, 2017), employment turnover following public outcry (Mourtgos et al., 2021), and other explanations situated in critical cultural theory (Cooper, 2003). In the case of Minneapolis, our analysis provides quantitative evidence that the proportion of Black people stopped was dramatically reduced while the proportion of 'Unknown' race greatly increased and that this change is correlated to the killing of George Floyd, as policing behaviors in months leading up to the killing remained stable even as COVID-19 measures were implemented across Minnesota. It has been suggested in other areas of the United States that law enforcement agencies purposefully misidentify the race of stopped individuals (Webster, 2021), although academic research on this is limited (Alwd 6th Ed.

Richardson et al., 2019). Routine police stop data is not sufficiently detailed to shed light on this. Future complementary research efforts could use interview techniques with city officials, police management, police officers, and affected citizens to provide a localized understanding of changes in police stop behavior (Nix & Wolfe, 2018).

Geographic visualization of person search stop rates for Black and White individuals across Minneapolis census tracts indicated differences in both the magnitude and location of policing. In areas with large amounts of police stops, such as Camden Hawthorne in northwest Minneapolis, person search rate for Black people far exceeds the person search rate for White people (Camden Hawthrone: person search rate 60.3 per 100 Black people, person search rate 6.6 per 100 White people). In addition, Black people are searched at very high rates in areas of Minneapolis that report high income and a large number of White residents, such as the East Isles and Calhoun areas. Given the descriptive nature of these visualizations, future research could apply numerous statistical techniques to further understand our observations. For example, a regression discontinuity or autoregressive integrated moving average approach could be applied to the temporal shifts in police stops. Additionally, several spatial statistical techniques could evaluate both clusters of police stops and the impact of area-level characteristics on police stop and search rates.

Though we were originally interested in Black people in particular, the results highlight the need to also look at the experience of Indigenous people interacting with police in Minneapolis. We found that the odds of being searched was drastically higher among Indigenous people as compared to White people, with an effect size higher than that estimated for Black people. This is consistent with recent research examining the policing of Indigenous peoples in Minneapolis (Gorsuch & Rho, 2019). A particularly interesting finding was that, among vehicle searches,

Indigenous people had higher odds of being searched *after* accounting for covariates, whereas across all previous models indicated that effect sizes decreased after adjustment. There are several possible explanations for this pattern. This could be due to imprecision in the model, as there were relatively few Indigenous people in the dataset. However, the confidence intervals are narrow, and as such it is also possible that these estimates are in fact accurate. Should this be the case, a finding such as this one could likely be explained by a systematic reason wherein Indigenous people are less likely to have their car searched due to a control variable included in our multivariate analysis. At this stage we do not know exactly why this might be the case, or which variable may be responsible, and further research into the topic should be done to understand the phenomenon.

While this study has many advantages, it is also important to mention its limitations. Certain variables considered in the regression analyses had missing data, specifically the pre-race information provided by dispatch. However, Black people still make up the largest proportion of stops by pre-race information, and these proportions are exacerbated when the looking at determined race. Thus, we argue that the amount of missing data is more likely explained by the difficulty of dispatchers providing pre-race information before stops occur.  In addition, stops for truancy and curfew violations were removed from the datasets because there were not enough data points. Although, we removed these observations from the final analysis, these observations comprised of a small percentage of all stops and so it is unlikely there would be any substantive impact on our conclusions.

While the open Minneapolis Police Stop Data provides detailed information on several important demographic and contextual factors, the data does not provide any follow-up information as to whether a stop leads to an arrest or the identification of contraband. The absence of this information prevents us from calculating and analyzing outcomes-based tests such as 'hit rates', a commonly used metric in research on police behavior (Persico & Todd, 2006; Shjarback et al., 2017). Hit rates describe the proportion of searches that lead to additional legal action. While providing another perspective of analysis, hit rates do not address differences in police behavior by race (Gelman et al., 2007) and are nested under a broader discussion on the difficulty identifying the appropriate baseline for the distribution of stopped individuals (S. Walker, 2008). In addition, the dataset does not contain any variables that have been shown to impact police-citizen interaction during stops, such as characteristics of the police officer, the type of police patrol, information on other individuals involved in the stop, or actions leading up to or taken during the stop (National Highway Traffic Safety Administration (NHTSA), 2020; Sykes & Clark, 1975; Warren et al., 2006). Demographic differences among these omitted variables may impact the estimated relationships between police stop outcomes by race in Minneapolis. Lastly, the Minneapolis Police Stop Data contain minimal information on what originally prompted the stop, which prohibits analyzing stops by minor and major infractions. The available variable describing what prompted the stop contains a large amount of missing data and the remaining variable values do not separate what prompted the stop by severity of infraction. Having additional information on the legal outcomes of stopped individuals, involved police officer, and information initiating the incident would provide further information on the behavior of the MPD regarding differences in police stop behavior.

## 5. Conclusion

Our study shows that certain populations, specifically Black and Indigenous people, are more likely to be stopped and searched by the Minneapolis Police Department (MPD) compared to White people. Temporal visualizations of police stops show dramatic changes in police stop activity immediately following the murder of George Floyd, with the number of stops and proportion of black people stopped reducing by half. Mapping of person search rates at the Census Tract-level show that Black people are searched at rates drastically higher compared to White people. These results corroborate previous work about racial bias in policing and show that, even when adjusting for confounding, Black and Indigenous populations experience a disproportionate burden of police intervention. This work contributes to a growing literature using open police data and multivariate statistical analyses to demonstrate clear statistical patterns in police behaviours along racial lines. Further analyses of such data sources like Minneapolis Open Data can provide insight to the pervasiveness of systemic racism in policing.

## Acknowledgements

E Rose is a postdoctoral fellow supported the National Institute of Mental Health of the National Institutes of Health under Award Number R01 MH114873. EEM Moodie acknowledges the support of a chercheur de mérite career award from the Fonds de Recherche du Québec - Santé and a Canada Research Chair (Tier 1). T Onookome-Okome, J Gorondonsky, J Sauer, and K Lum received no external funding for participating in this research.

**Tables**

Table 1: Demographic breakdown of Minneapolis population compared to MPD police stops

| Race | Proportion of Minneapolis Population[1] | Proportion of MPD Police Stops |
|---|---|---|
| White | 63.6 | 20.9 |
| Black | 19.2 | 34.7 |
| Asian | 5.9 | 1.2 |
| Hispanic or Latino | 9.6 | 3.4 |
| Indigenous | 1.4 | 2.6 |
| NA/Unknown/Missing | n/a | 35.7 |

1. Data retrieved from the U.S. Census Bureau QuickFacts statistics system. Underlying data from the Population Estimates Program (PEP) and American Community Survey (ACS). Margin of error and differences in racial/ethnic categorization does not accumulate to 100%. Available from: https://www.census.gov/quickfacts/minneapoliscityminnesota

Table 2. Summary statistics for stop data by stop outcome (person search, vehicle search) and individual characteristics. Percentages are additive for a given outcome category.

| Characteristic | Person search | | Vehicle search | |
| --- | --- | --- | --- | --- |
| | **Yes** | **No** | **Yes** | **No** |
| | **N (%)** | **N (%)** | **N (%)** | **N (%)** |
| *Race information provided by dispatch* | | | | |
| Asian | 75 (0.4%) | 355 (0.3%) | 31 (0.3%) | 399 (0.3%) |
| Black | 6267 (32.6%) | 18562 (13.9%) | 2485 (20.6%) | 22344 (15.9%) |
| Latino | 270 (1.4%) | 1497 (1.1%) | 99 (0.8%) | 1668 (1.2%) |
| Indigenous | 766 (4%) | 2047 (1.5%) | 141 (1.2%) | 2672 (1.9%) |
| Other | 235 (1.2%) | 1200 (0.9%) | 88 (0.7%) | 1347 (1%) |
| Unknown | 9739 (50.7%) | 94746 (71.1%) | 8625 (71.5%) | 95860 (68.3%) |
| White | 1872 (9.7%) | 14770 (11.1%) | 598 (5%) | 16044 (11.4%) |
| *Determined Race* | | | | |
| Asian | 165 (0.9%) | 1847 (1.4%) | 110 (0.9%) | 1902 (1.4%) |
| Black | 12603 (65.6%) | 47512 (35.7%) | 8013 (66.4%) | 52102 (37.1%) |
| Latino | 731 (3.8%) | 5222 (3.9%) | 482 (4%) | 5471 (3.9%) |
| Indigenous | 1444 (7.5%) | 2994 (2.2%) | 506 (4.2%) | 3932 (2.8%) |
| Other | 485 (2.5%) | 3243 (2.4%) | 261 (2.2%) | 3467 (2.5%) |
| Unknown | 536 (2.8%) | 39464 (29.6%) | 1075 (8.9%) | 38925 (27.7%) |
| White | 3260 (17%) | 32895 (24.7%) | 1620 (13.4%) | 34535 (24.6%) |
| *Reported Problem* | | | | |
| Attempted Pick-Up | 795 (4.1%) | 1844 (1.4%) | 49 (0.4%) | 2590 (1.8%) |
| Suspicious Person | 7304 (38.0%) | 33043 (24.8%) | 678 (5.6%) | 39669 (28.3%) |
| Suspicious Vehicle | 3606 (18.8%) | 31292 (23.5%) | 3747 (31.1%) | 31151 (22.2%) |
| Traffic Law Enforcement | 7519 (39.1%) | 66998 (50.3%) | 7593 (62.9%) | 66924 (47.7%) |
| *Gender* | | | | |
| Female | 2774 (14.4%) | 28605 (21.5%) | 1583 (13.1%) | 29796 (21.2%) |
| Gender Non-Conforming | 27 (0.1%) | 217 (0.2%) | 16 (0.1%) | 228 (0.2%) |
| Male | 16179 (84.2%) | 76850 (57.7%) | 9655 (80%) | 83374 (59.4%) |
| Unknown | 244 (1.3%) | 27505 (20.7%) | 813 (6.7%) | 26936 (19.2%) |

Notes: Describes all stops with data from Minneapolis Open Police Dataset from October 31[st], 2016, to March 15[th], 2021.

Table 3. GEE Models: Estimated odds ratios and 95% confidence intervals for association between race and person/vehicle search among individuals who were stopped by the police in Minneapolis, MN before/after the murder of George Floyd compared to White people. Multivariate models control for gender, pre-race coding, problem, reason, whether the stop was after sunset, latitude, and longitude. * indicates odds ratios that are significantly different than 1 with the false discovery rate controlled at 0.05 using the Benjamini-Hochberg procedure.

|  | Outcome: Person Search | | Outcome: Vehicle Search | |
| --- | --- | --- | --- | --- |
|  | Univariate | Multivariate | Univariate | Multivariate |
| Black | 2.01* (1.59, 2.53) | 1.72* (1.54, 1.92) | 2.26* (2.08, 2.45) | 2.02* (1.83, 2.24) |
| Latino | 1.27* (1.11, 1.45) | 1.26* (1.09, 1.46) | 1.64* (1.43, 1.88) | 1.49* (1.25, 1.78) |
| Indigenous | 3.96* (3.35, 4.68) | 3.52* (3.01, 4.11) | 2.21* (1.88, 2.60) | 5.36* (4.37, 6.56) |
| Asian | 0.80* (0.65, 0.97) | 0.81* (0.68, 0.97) | 0.99 (0.77, 1.28) | 0.83 (0.62, 1.10) |
| Other | 1.32* (1.12, 1.56) | 1.34* (1.18, 1.52) | 1.37* (1.18, 1.58) | 1.37* (1.19, 1.58) |
| Unknown | 0.08* (0.06, 0.11) | 0.17* (0.14, 0.21) | 0.39* (0.34, 0.45) | 0.62* (0.51, 0.75) |

Table 4. GEE Models: Estimated odds ratios and 95% confidence intervals for association between race and person/vehicle search among individuals who were stopped by the police in Minneapolis, MN after the murder of George Floyd compared to White people. Multivariate models control for gender, pre-race coding, problem, reason, whether the stop was after sunset, latitude, and longitude. * indicates odds ratios that are significantly different than 1 with the false discovery rate controlled at 0.05 using the Benjamini-Hochberg procedure.

|  | Outcome: Person Search | | Outcome: Vehicle Search | |
| --- | --- | --- | --- | --- |
|  | Univariate | Multivariate | Univariate | Multivariate |
| Black | 2.77* (2.07, 3.71) | 2.37* (1.94, 2.90) | 3.11* (2.45, 3.95) | 2.55* (1.92, 3.37) |
| Latino | 1.98* (1.37, 2.87) | 1.82* (1.29, 2.58) | 1.81* (1.17, 2.78) | 1.56 (0.92, 2.62) |
| Indigenous | 4.59* (3.08, 6.85) | 4.67* (3.30, 6.61) | 2.37* (1.73, 3.25) | 3.30* (2.26, 4.83) |
| Asian | 1.03 (0.50, 2.09) | 1.02 (0.55, 1.88) | 1.40 (0.65, 3.01) | 1.25 (0.54, 2.91) |
| Other | 0.99 (0.50, 1.95) | 1.16 (0.64, 2.08) | 1.80 (0.93, 3.47) | 1.59 (0.76, 3.34) |
| Unknown | 0.04* (0.02, 0.08) | 0.13* (0.09, 0.20) | 0.37* (0.28, 0.49) | 0.57* (0.49, 0.82) |

**Figures**

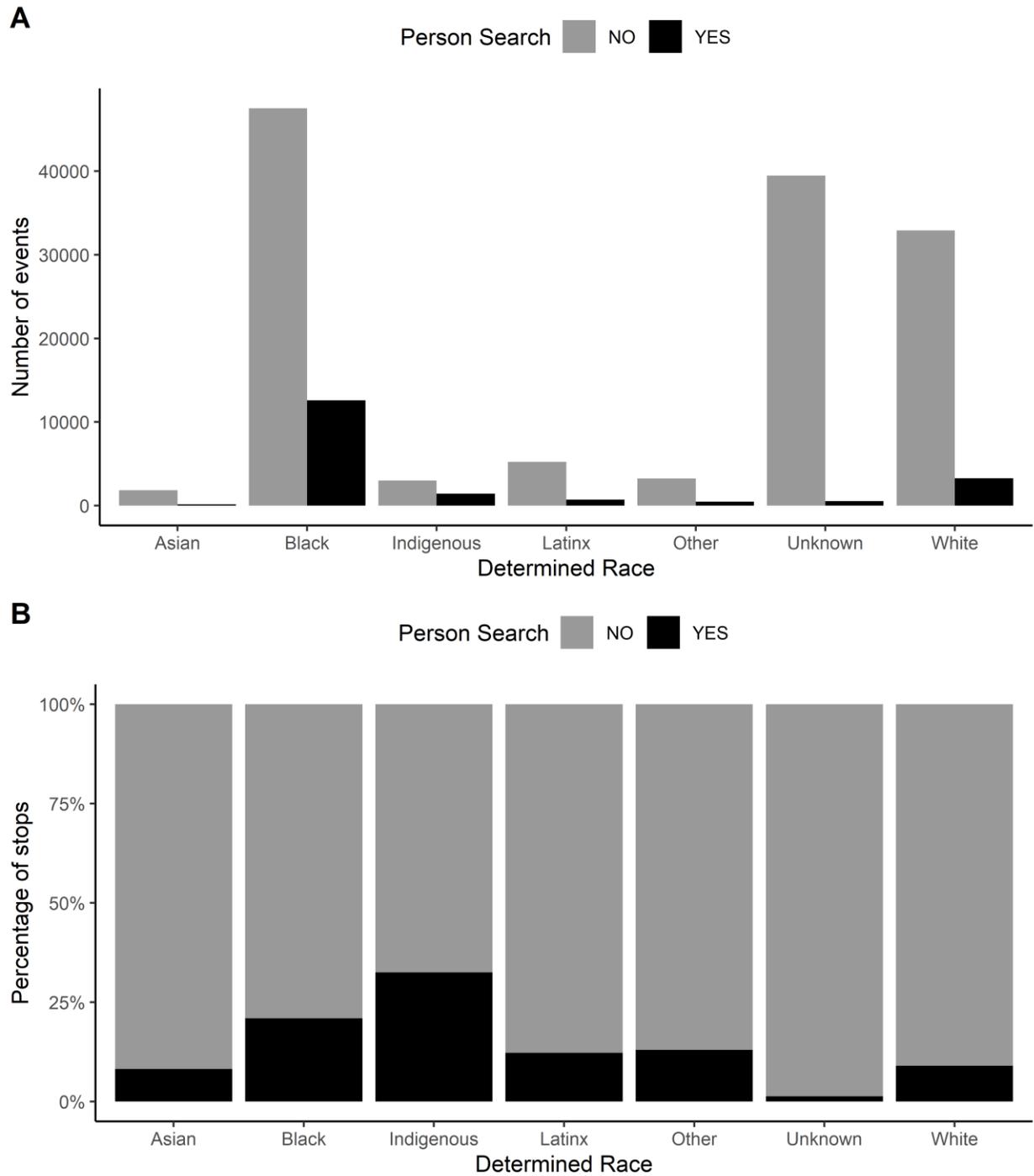

Figure 1. **A:** Absolute number of police stops resulting in search by race. **B:** Percentage of stops ending in search by race. Describes all stops with data from Minneapolis Police Stop Data from October 31st, 2016, to March 15th, 2021.

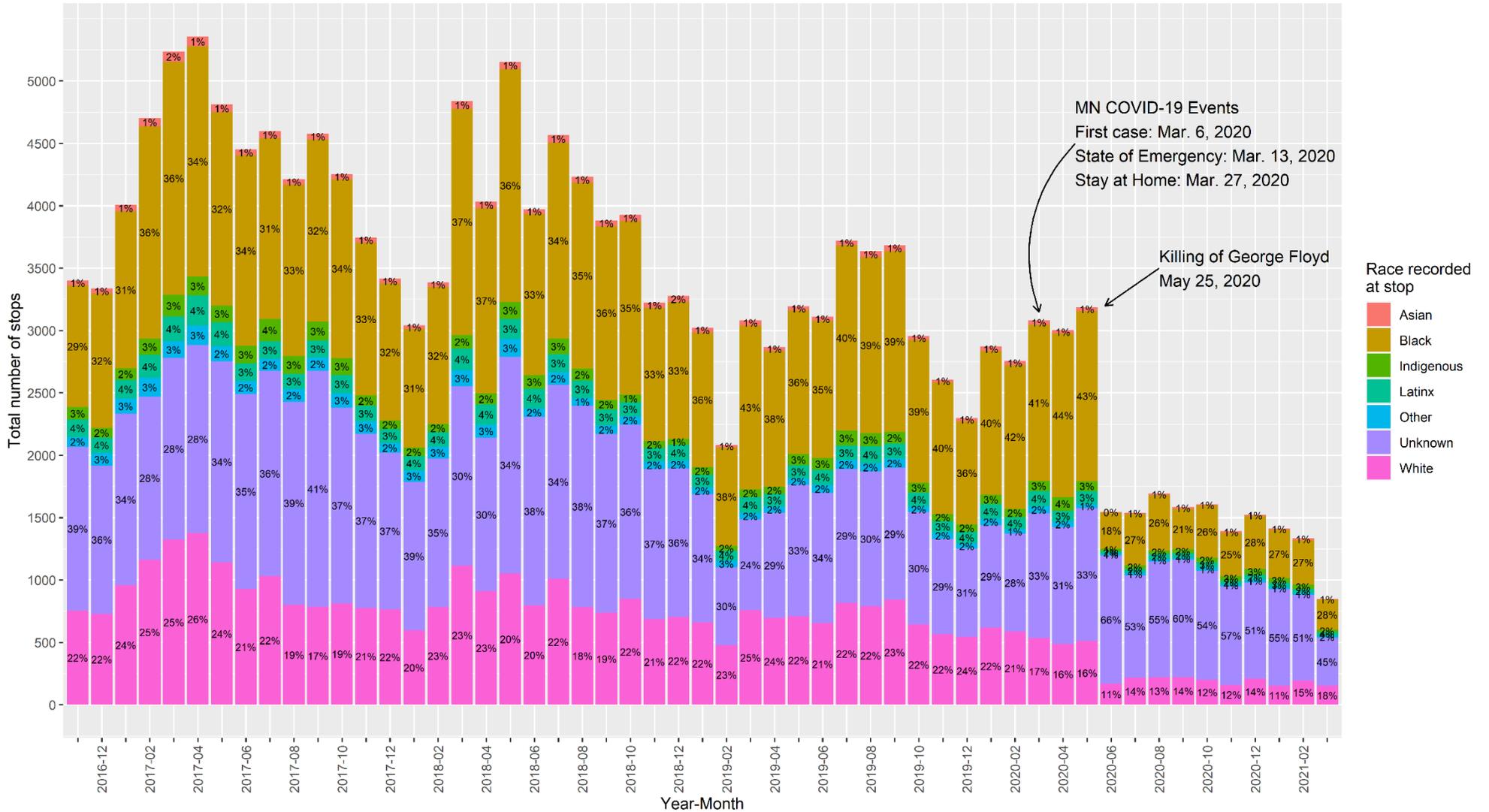

Figure 2. Time-series bar chart of Minneapolis Police Department stops by race. Timespan ranges from November 2016 to mid-March 2021. Each bar describes the distribution of police stops by race in a month-year.

### A) Person search stop rate for Black individuals, 2017

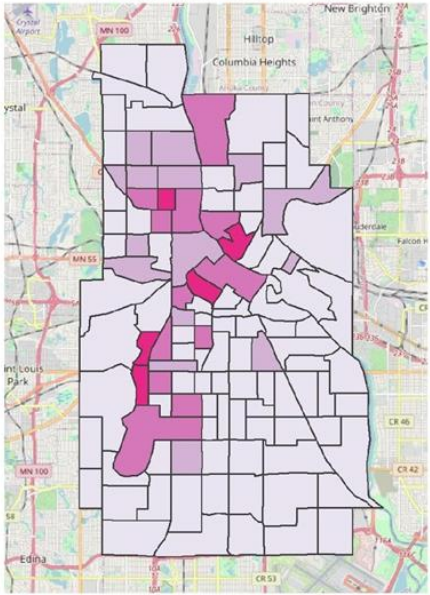

### B) Person search stop rate for Black individuals, 2018

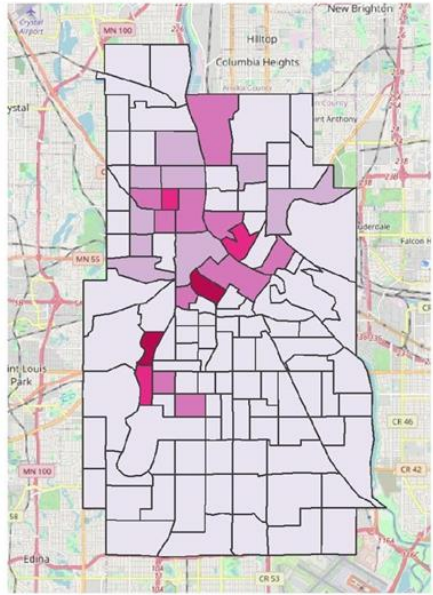

### C) Person search stop rate for Black individuals, 2019

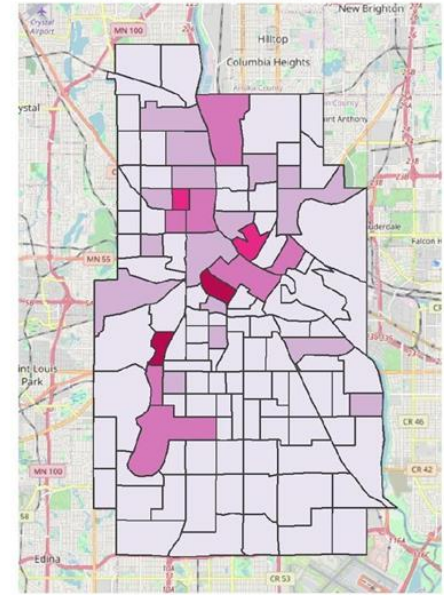

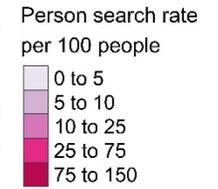

Person search rate per 100 people

- 0 to 5
- 5 to 10
- 10 to 25
- 25 to 75
- 75 to 150

### D) Person search stop rate for White individuals, 2017

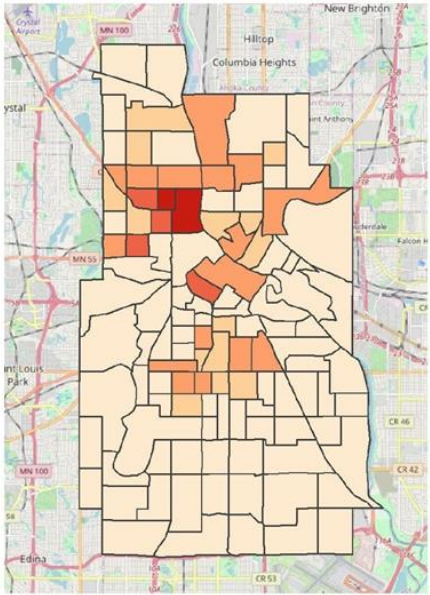

### E) Person search stop rate for White individuals, 2018

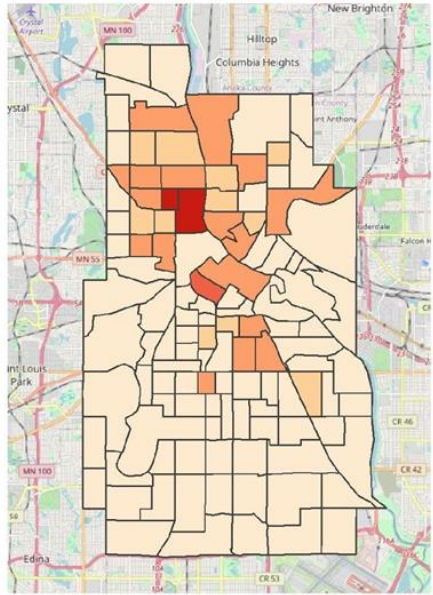

### F) Person search stop rate for White individuals, 2019

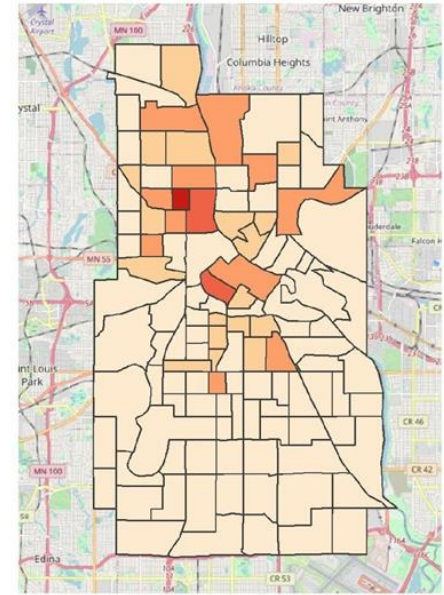

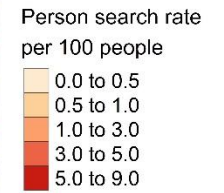

Person search rate per 100 people

- 0.0 to 0.5
- 0.5 to 1.0
- 1.0 to 3.0
- 3.0 to 5.0
- 5.0 to 9.0

Figure 3. Choropleth maps of person search stop rates across Minneapolis at the Census Tract level. Maps are separated by year and Black/White race. Note that the scales used between the two maps differ. Rates are calculated as the number of person searches for a given race divided by the population of that race in the Census Tract. Population data from the U.S. Census Bureau American Community Survey. Stop data from the Minneapolis Police Department.

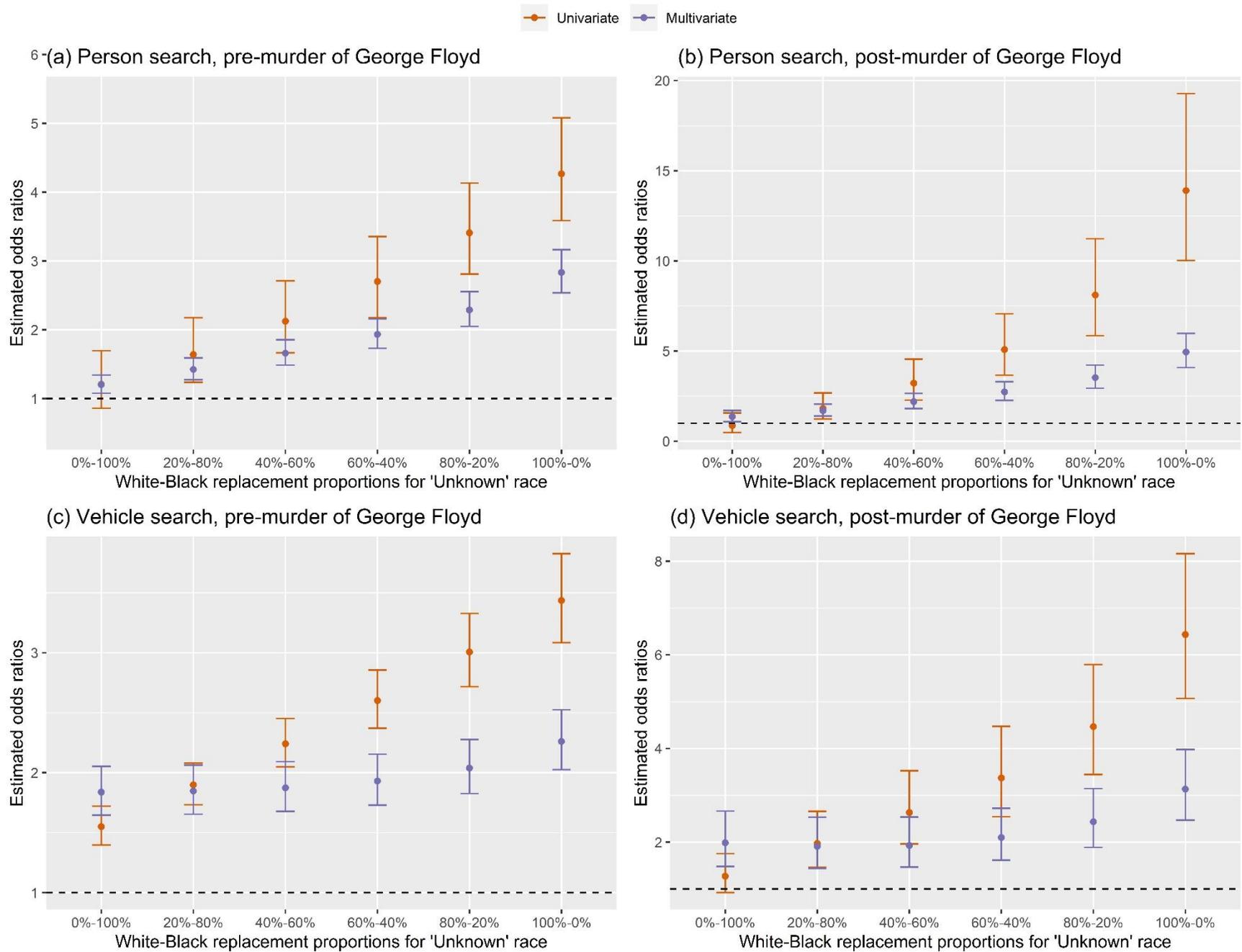

Figure 4. Comparison of estimated odds ratios and 95% confidence intervals when replacing the 'Unknown' race category with different proportions of white and Black individuals. Only the odds ratios for Black individuals are shown, with a referent category of white individuals. Replacement race assignment to white or Black is random.